\documentclass{article}

\usepackage{microtype}
\usepackage{graphicx}
\usepackage{subfigure}
\usepackage{booktabs} 

\usepackage{hyperref}

\PassOptionsToPackage{numbers,authoryear}{natbib}
\usepackage[preprint]{neurips_data_2024}

\usepackage{amsmath}
\usepackage{amssymb}
\usepackage{mathtools}
\usepackage{amsthm}

\usepackage[]{natbib}
\theoremstyle{plain}

\theoremstyle{definition}

\theoremstyle{remark}

\usepackage[textsize=tiny]{todonotes}

\begin{document}

\title{Using Galaxy Evolution as Source of Physics-Based Ground Truth for Generative Models}

\author{%
  Yun Qi Li  \\
  Physics and Astronomy Department\\
  UCLA\\
  Los Angeles, CA 90095\\
  \texttt{yunqil@g.ucla.edu} \\  
  \And
  Tuan Do\\
  Physics and Astronomy Department\\
  UCLA\\
  Los Angeles, CA 90095\\
  \texttt{tdo@astro.ucla.edu} \\
  \And
  Evan Jones  \\
  Physics and Astronomy Department\\
  UCLA\\
  Los Angeles, CA 90095\\
  \And
  Bernie Boscoe \\
  Computer Science Department \\
  Southern Oregon University \\
  Ashland, OR 97520 \\
  \And
  Kevin Alfaro \\
  Physics and Astronomy Department\\
  UCLA\\
  Los Angeles, CA 90095\\
  \And
  Zooey Nguyen \\
  Physics and Astronomy Department\\
  UCLA\\
  Los Angeles, CA 90095\\
}

\maketitle

\begin{abstract}

Generative models producing images have enormous potential to advance discoveries across scientific fields and require metrics capable of quantifying the high dimensional output. We propose that astrophysics data, such as galaxy images, can test generative models with additional physics-motivated ground truths in addition to human judgment. 
For example, galaxies in the Universe form and change over billions of years, following physical laws and relationships that are both easy to characterize and difficult to encode in generative models. 
We build a conditional denoising diffusion probabilistic model (DDPM) and a conditional variational autoencoder (CVAE) and test their ability to generate realistic galaxies conditioned on their redshifts (galaxy ages). 
This is one of the first studies to probe these generative models using physically motivated metrics. 
We find that both models produce comparable realistic galaxies based on human evaluation, but
our physics-based metrics are better able to discern the strengths and weaknesses of the generative models. 
Overall, the DDPM model performs better than the CVAE on the majority of the physics-based metrics. 
Ultimately, if we can show that generative models can learn the physics of galaxy evolution, they have the potential to unlock new astrophysical discoveries. 

\end{abstract}

\section{Introduction} \label{Introduction}

The intersection of high dimensional data and machine learning in fields such as particle physics, genomics, and astrophysics has seen huge advances in the ability to extract meaningful information and patterns from complex data. Image data, in particular, is one example of data with large numbers of features. Advances in machine learning have high potential to advance scientific discovery in astrophysics, which amasses large amounts of high dimensional data including multi-wavelength images, spectra, and time series ($>10^9$ objects forthcoming in large-scale surveys over the next decade). Generative models have risen to the fore to study relationships among features and statistical structures of distributions behind audio, image or video data. 

Generative models have been very successful at producing real-world images that are termed `realistic' based on human perception. Major methods include variational autoencoders (VAE) \citep{kingma2013}, generative adversarial networks (GAN) \citep{goodfellow2014}, and denoising diffusion probabilistic models (DDPM) / denoising diffusion implicit models (DDIM) \citep{sohl-dickstein2015,ho2020,song2020}. The combination of human preference and judgement is a common approach to evaluate images produced by generative models. Humans are very efficient at identifying irregularities in artificially generated images, such as a missing eyeglass frame in the image of a person or an image of spiral galaxy with a malformed arm. However, examining millions of generated images manually is an impractical task for humans. 

To further human-like scoring methods, popular metrics such as the Inceptiuon Score (IS) \citep{salimans2016} or the Frechet Inception Distance (FID) \citep{heusel2018} were developed to produce scores that correlate well with human judgement. The IS focuses on measuring the diversity of the generated samples and whether the generated images represent clearly defined objects \citep{salimans2016}. The FID improves upon the IS by incorporating a ground truth, and comparing the generated dataset with the real dataset in a feature space \citep{heusel2018}. Both metrics are highly successful in providing a quantitative and efficient measure of human perception of image quality and have played major roles in the development and improvement of generative models. However, as artificially generated images become indistinguishable to humans from real images, metrics based on human perception become less useful. 

\begin{figure*}[thb]
\centering
    \includegraphics[width=5in]{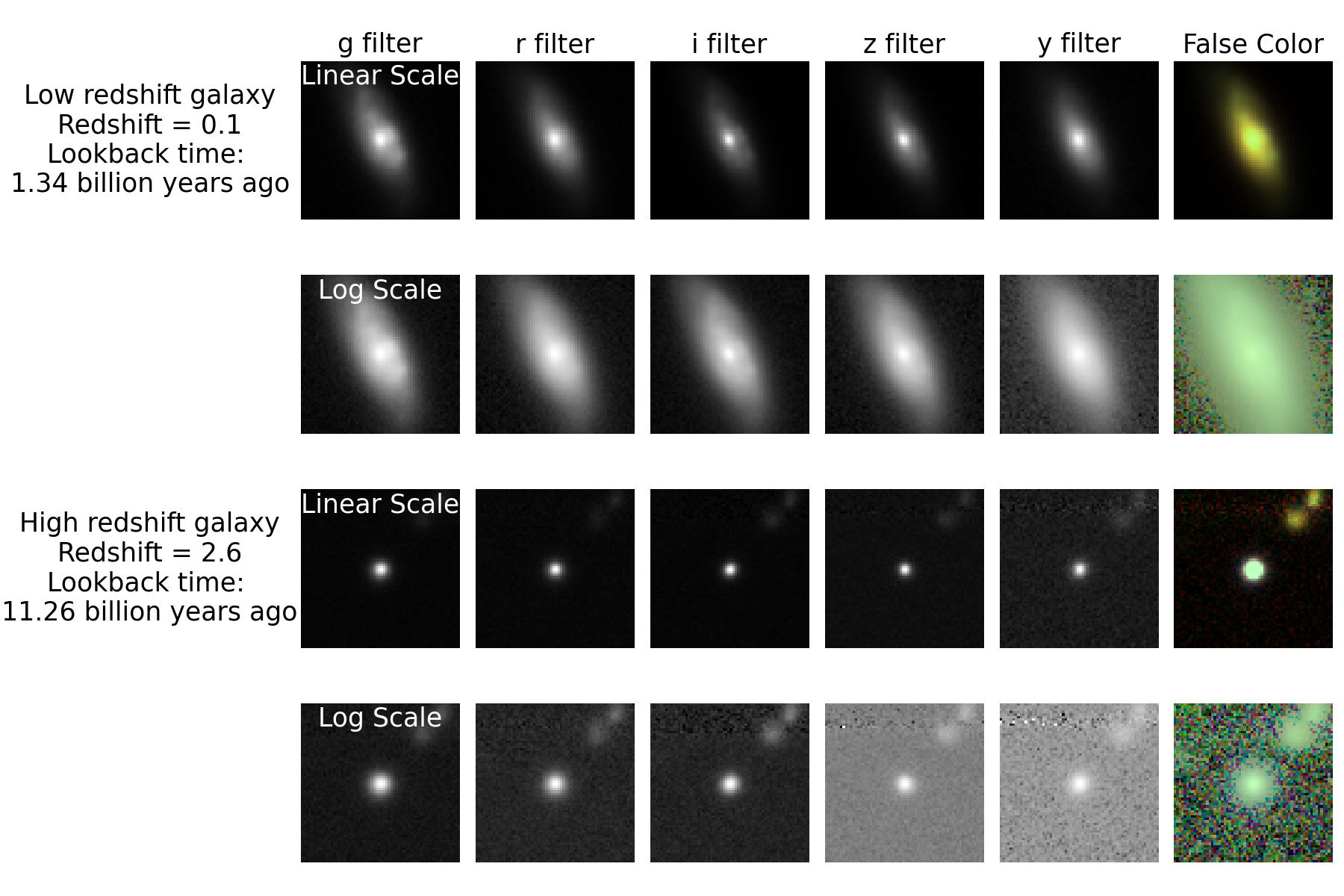}
    \caption{Grid of images showing two example galaxies in our training dataset. The first two rows shows a low redshift galaxy (closer to Earth) while the last two rows show a higher redshift galaxy (farther from Earth). Each galaxy is observed using 5 optical filters represented has the first five columns. The false color image using these five filters is shown in the last column. We show the images both with linear and log scaling to show different features. Note that in general galaxies closer to Earth have larger sizes and more complex features while those further away are more compact in appearance. }
    \label{fig:real_galaxies_showcase}
\end{figure*}

We propose that additional physics-based metrics can be applied to astrophysical datasets of galaxy images as additional sources for ground truth to improve and generalize image generation models. To successfully generate images of galaxies, a range of features must be reproduced, including shapes and sizes, a dynamic range in brightness (often $>10^6$ contrast between the lowest and highest pixel values) and properties relating to the evolution of galaxies over time. The most important factor that governs the appearance of galaxies is their redshift, which corresponds to both the distance from the object to Earth as well as the time span between the moment at which the light from the galaxy was emitted and observed by us (lookback time). Human-based perception metrics can be used to evaluate the quality of generated galaxy images, but those metrics may miss scientifically important features. For example, the distribution of sizes of galaxies should follow a certain shape for galaxies at a given redshift. This is because there is a physical relationship between the appearance of a galaxy and its age \citep{robertson2019}. Features of galaxies are parametrizable, giving rise to the possibility of directly measuring and evaluating their quality. Such measurements can be done using common tools in astronomy like Source Extractor \citep{bertin1996}.

In this work, we build a conditional variational autoencoder (CVAE) \citep{sohn2015} and a conditional denoising diffusion probabilistic model (DDPM) \citep{ho2020} to generate images of galaxies conditioned on their respective redshifts. Our models were trained using a dataset of $\sim300$k galaxies \citep{do2024,jones2024} with redshifts between 0.1 and 4 (corresponding to a lookback time of 1.54 billion and 13.6 billion years). We propose metrics that are closely tied to the physical properties of galaxies in addition to human-perception based IS and FID. These metrics include the galaxy-fitting loss, the galaxy KL loss, and the redshift loss. Finally, we compare the images generated by these two models using these metrics and discuss relative strengths and weaknesses as a proof of concept for using this approach to evaluating generative models.

\section{Related Works}


In recent years, astronomers have been testing generative models such as CVAE and DDPM to evaluate their ability to produce galaxy images with good pixel statistics and visual quality \citep{fussell2019,lanusse2021,smith2022}. We expand this approach by focusing on the physics governing the appearance of these galaxies over time using generative models.

\citet{ravanbakhsh2017} employed conditional variational autoencoders (CVAEs) and conditional generative adversarial networks (CGANs) to simulate images from the GALAXY-ZOO and Hubble Space Telescope Advanced Camera for Surveys (HST/ACS) COSMOS survey. They conditioned these models on parameters including the half-light radius, magnitude, and redshift of galaxies, allowing the models to focus on the visual quality of the generated images. The generated dataset is evaluated using pixel statistics. This study found that CVAEs are useful tools to denoise galaxy images. 

\citet{fussell2019} advanced this approach by simulating galaxy image data using chained generative adversarial networks, a technique inspired by the StackGAN text-to-image model \citep{zhang2016}. They evaluated their generated datasets by comparing the distributions of the physical properties of the galaxies. Differing from the approach of \citet{ravanbakhsh2017}, \citet{fussell2019}'s models are not conditioned on physical parameters. They conclude that state-of-the-art generative models are useful for astronomy, but they did not examine the relationships of the generated features or as a function of the age of the galaxy. 

\citet{lanusse2021} developed variational autoencoders for simulating images of galaxies selected from the HST/ACS COSMOS survey and evaluating the generated galaxies based on their physical properties. They evaluated the generative models in terms of the statistical relationship between the features, such as between the ellipticity of galaxies and their brightness. 

More recently, \citet{smith2022} utilized denoising diffusion probabilistic models to simulate galaxy images, focusing on the visual qualities of the generated galaxies. Using DESI as the dataset source, they selected galaxies that are bright and nearby (redshift $<0.08$) with clearly defined galaxy features for each image. Their model generated galaxy images that are visually indistinguishable from real galaxy images based on human evaluation. \citet{smith2022} evaluated their generated dataset using the FID metric. They also explore DDPM's ability to regenerate images, through a full forward and reverse process. They proposed the Synthetic Galaxy Distance metric, which quantifies the difference between the regenerated image and the original image based on the galaxy's size and brightness.

Recently, \citet{yin2022} explored using conditional autoencoders conditioned on the mean value of the real image to generate new instances of the galaxy image. Their goal is to segment the original input galaxy images through passing them through the CAE, and interpret galaxy parameters from the latent space of the CAE model. They evaluated their models based on comparing the latent space extracted parameters to the measured values. 

Our study builds on previous approaches by exploring machine learning methods capable of reproducing galaxies with both accurate physical properties and high visual fidelity, conditioned solely on the redshift (i.e. lookback time) of the galaxy. Since our purpose is to generate images with the same diversity properties as seen in real galaxies, we do not attempt to remove complexity or diversity from the dataset through data processing. We also do not provide the models any physical parameter that it has to learn. The assessment of the models are based on their ability to generate galaxies with characteristics such as the size and brightness of galaxies, at varying redshifts. By evaluating their capacity to understand the physics of galaxies, we explore their potential to learn and reproduce deep relationships in images. Such methods can be used to improve generative models in the long run. 

\subsection{Evaluation Metrics for Generative Models}

It is a challenge to quantitatively evaluate generated images because patterns in images are often subject to deep underlying relationships that are not quantifiable with pixel-based statistics. Previous research has produced metrics that correlate well with human perception. These metrics include the IS (IS) \citep{salimans2016} and the 
Frechet Inception Distance (FID) \citep{heusel2018}. However, these metrics become less indicative when selecting generative methods for scientific use. Moreover, for cases where humans cannot assess the quality of generated images, or distinguish generated images from real ones, it is questionable to use a metric that corresponds well with human judgement. To address this gap in evaluation methods, we propose our metrics based on galaxy image data.
In this study, we use galaxy images as a new form of ground truth in generative models of images. Galaxies possess deep patterns just like ordinary objects. Some of these patterns, such as the relationship between the flux and redshift, are difficult to perceive even for humans. Conversely, galaxies can be parametrized for their physical properties and quantitatively evaluated. Improving the physics of generated galaxies is more objective than metrics of human perception, and therefore, generated galaxies based metrics are more useful for scientific use. 

\section{Methods}

\subsection{Metrics}

We develop metrics to compare the distribution of physical properties of galaxies generated by our models to those of real images to assess how physically realistic the generated images are. We measure galaxy features from both the generated and real images in the same way, then compare the distribution of these properties using the KL divergence and what we term the galaxy fitting loss. These galaxy properties are chosen to represent physically meaningful features of galaxy evolution. Galaxies today have a larger range of sizes and shapes than galaxies from long ago \citep{conselice2014}. In addition, we also put the generated galaxy images through a trained CNN model that can predict galaxy redshifts to examine how well the generated image has been conditioned on redshifts. We use this comparison to create a redshift loss. We describe these metrics in detail below.

\begin{figure}[thb]
\centering
    \includegraphics[width=5.0in]{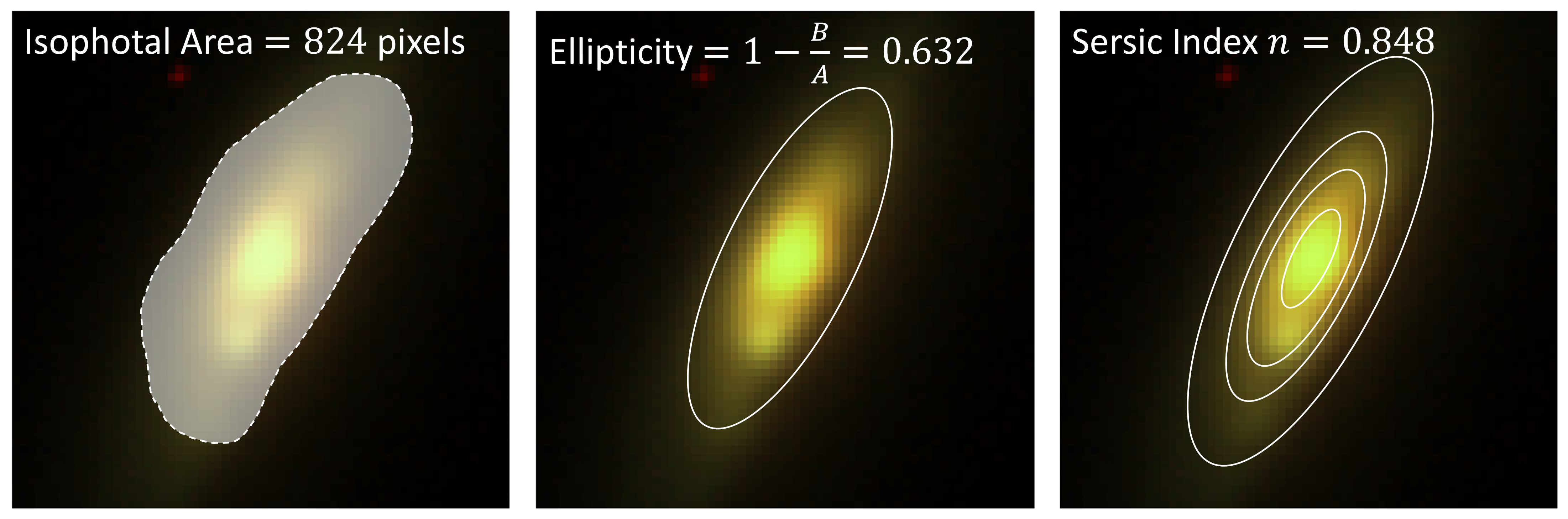}
    \caption{Illustration of the three galaxy metrics that we measure from the images. Left: isophotal area. The highlighted pixels indicate area of the image that exceed the background threshold. Middle: Ellipticity. The ellipitical frame shows the ellipticity of the best fit Gaussian profile. Right: Sersic index. The contours show how brightness is distributed with relation to radius. }
    \label{fig:galaxy_metrics}
\end{figure}

\subsection{Fitting Individual Galaxies}

Source Extractor, a common astrophysical image analysis tool, is used to measure features of galaxies connected the physics of their formation and evolution \citep{bertin1996}. Source Extractor first identifies sources in the image using a 2D Gaussian fit and detects sources above $>3 \sigma$ above the background RMS pixel values to produce an image segmentation map. In this work, we concentrate on three parameters that describe the shape and flux distribution of the central galaxy: isophotal area, ellipticity, and Sersic index.
Isophotal area is the number of pixels with values exceeding a threshold above the background ($>1.5$ $\sigma$). The ellipticity measures its shape in terms of a fit to an ellipse $e = 1-(a/b)$, where $a$ is the semi-major axis and $b$ is the semi-minor axis. 
Sersic profile measures how the intensity ($I$, flux per unit area) is distributed in the galaxy as a function of projected radius from the galaxy center:
\begin{equation}
I(R)=I_e \exp\left\{ -b_n\left[ \left( \frac{R}{R_e}\right) ^{1/n} -1\right] \right\},
\end{equation}
where $R_e$ is the half-light radius, $I_e$ is the intensity at that radius, and $n$ is the Sersic index that measures how concentrated the light is at the center (larger values of $n$ means more centrally concentrated).

\subsection{KL Divergence}
The first metric, the galaxy Kullback-Leibler (KL) loss, quantitatively compares the distribution of physical parameters from the generated galaxies to those from real images. We use the Source Extractor output for each of isophotal area, ellipticity, and Sersic index to obtain the probability distribution for each parameter at different redshift, and calculate the KL divergence between these distributions. Since these parameters evolve with redshift (time), we split each distributions into 10 bins between the minimum and the maximum value to calculate the KL divergence. The galaxy KL loss is the average for the KL divergence across all parameters and bins. 
\begin{equation}
\text{KL Loss}= \frac{1}{N_{\text{params}}} \sum_{i=1}^{N_{\text{params}}} \left( \frac{1}{N_{\text{bins}}} \sum_{j=1}^{N_{\text{bins}}} \text{KL} \left( P_{\text{real}, i, j}, P_{\text{gen}, i, j} \right) \right)
\end{equation}
In practice, we find it is more meaningful to compare the ratio of the KL divergence between each model and the real data. 

\subsection{Galaxy-fitting Loss}
The second metric galaxy-fitting loss aims to measure the irregularity of the generated galaxies. 
Source Extractor code we use for fitting model intensity distributions to the images is able to also estimate uncertainties in the model parameters as well as the fit residuals. We find that such uncertainties correlate well with the noise-to-signal ratio of the image, or the irregularity of the generated galaxy. A combined uncertainty is calculated within Source Extractor (internally called the ERRCXY\_IMAGE parameter). 
We define the galaxy-fitting loss the mean $\log$ of the "ERRCXY\_IMAGE" values. The choice to take the mean after the log is to reduce the effects of outlier values:
\begin{equation}
\text{Galaxy Fitting Loss} = \frac{1}{N} \sum_{i=1}^{N} \log(\text{`ERRCXY\_IMAGE'}_i)
\end{equation}
We use the noise residuals from fitting real galaxies as a baseline for the expected loss. 

\subsection{Redshift Loss}

We define the redshift loss as the difference between the redshift the generated models are conditioned on, compared to the redshift encoded in the generated images.
\begin{equation}
\text{Redshift Loss} = \frac{1}{N} \sum_{i=1}^{N} \left( 1 - \frac{1}{1 + \left( \frac{(z_{\text{pred}, i} - z_{\text{cond}, i})}{0.15} \right)^2} \right)
\end{equation}
Where $z_{pred}$ is the CNN predicted redshift (the redshift encoded in the images), and $z_{cond}$ is true redshift (the redshift conditioned at time of image generation). 
The form of this loss is defined in \citet{nishizawa2020}. In this case, the redshift discriminator measures the difference between the redshift the model is asked to produce and the redshift it actually recovered in the generated images. The redshift loss is compared across the range of redshifts for model generated images and the real dataset.

Redshift loss quantifies the generative models’ ability to recover the relationship between the redshift of galaxies and their appearance. To compute redshift loss, we use a pre-trained convolutional neural network (CNN) redshift discriminator (Jones et al. in prep.). This approach is similar to the IS, where a CNN is employed for the metric. However, since the predicted redshift is a physically motivated value value, we can directly compare and measure the bias, scatter, and outlier rate between the conditioned values and the actual values encoded. 

\subsection{IS and Fréchet Inception Distance}

We compute the IS (IS) and the Fréchet Inception Distance (FID) using the definitions from \citet{salimans2016} and \citet{heusel2018} (see Appendix \ref{appendix:scores}). Both metrics require feeding the data into an Inception V3 model with the input of 3-channel RGB images with size of $299\times299$ pixels. To use the galaxy images, which consists of 5 filters, we create five 3-channel images by duplicating each filter into the 3 channels. Each image is super-sampled from $64\times64$ pixels to $299\times299$ pixels using the ''cv2.INTER\_LINEAR" function. Then, we compute the IS and FID for each of the five images and average the results. We compute the FID for the testing dataset and both generated datasets against the validation dataset ($2,000$ real galaxy images). 

\subsection{Datasets: Galaxy Images}

We adopt our dataset from \citet{do2024}, based on the second data release of the Hyper Suprime-Cam survey \citep{aihara2019}. The Hyper Suprime-Cam survey is a deep sky survey covering a wide range of galaxy (from those observed today to galaxies from 12 billion years ago), allowing us to explore galaxy evolution relationships in a large timescale. The data also only encompasses images in the visible and near-infrared band, increasing the challenge to correctly condition the redshift information. At last, the dataset we use is not truncated or normalized, in order to present the full range of physical properties to the model. 

The dataset consists of $286,401$ galaxies spanning redshifts between 0 and 4, with each galaxy captured with images in five visible light bands ($g,r,i,z,y$ filters) and represented with $64\times64$ pixels. Each galaxy has a true redshift (age) label assigned to it, from accurate spectroscopic measurements. Due to bias in the selection process, our dataset is skewed toward galaxies at lower redshifts (with $92.8$ percent of galaxies with redshift $< 1.5$). This poses an additional challenge for the generative models - the model is likely to produce worse images for galaxies beyond redshift of 1.5. This dataset is available in \citet{do2024} at \url{https://zenodo.org/records/11117528}.

The training dataset that we train the generative models on consists of $204,513$ galaxies, while the testing dataset for evaluating galaxy metrics (KL divergence, galaxy-fitting loss, and redshift loss) consists of $40914$ galaxies. The IS and FID for the real dataset are evaluated on a subset of the testing dataset with $2,000$ galaxies distributed uniformly across the bins shown in Figure \ref{fig:galaxy_kl_loss}. We also sample a validation dataset of $2,000$ galaxies with uniform redshift as the ground truth for the FID metric from the remaining $40914$ galaxies. 

\subsection{Models} \label{models}

\subsubsection{Conditional Denoising Diffusion Probabilistic Model} \label{ddpm}

To create a conditional DDPM, we modified and added a condition mechanism to the DDPM developed by \citet{ho2020}. We include the conditional component using an approach similar to \citet{mirza2014}, we reshaped redshift input to match the input dimensions of each down-sampling and upsampling blocks of the U-Net. This reshaped hidden representation is concatenated with the feature map input of each down-sampling and upsampling blocks. The forward process follows the non-conditional $x_t = \sqrt{1 - \beta_t}x_{t-1} + \sqrt{\beta_t}\varepsilon_{t-1}; \text{ where } \varepsilon_{t-1} \sim \mathcal{N}(0, I)$. $\beta_t$ is the series of hyperparameters defining the strength of noise added at each step \citep{ho2020}. We sample during the reverse process from a modified distribution, which the model predicts the added noise based on the additional redshift input $z$ \citep{lee2022}:
\begin{equation}
    p(x_{t-1}) = \frac{1}{\sqrt{\alpha_t}} \left( x_t - \frac{1-\alpha_t}{\sqrt{1-\overline{\alpha_t}}} \varepsilon_\theta(x_t, t, z) \right) + N(0, \sigma_t),
\end{equation}
where we assume $\sigma_t^2=\beta_t$. $\alpha_t$ and $\overline{\alpha_t}$ are series precomputed from the noise schedule $\beta_t$ \citep{sohl-dickstein2015,ho2020}. $\varepsilon_\theta(x_t, t, z)$ is the DDPM model predicted value of $\varepsilon_{t}$ sampled during the forward process. During training, we minimize the mean squared error between $\varepsilon_\theta(x_t, t, z)$ and $\varepsilon_{t}$. $\varepsilon_\theta(x_t, t, z)$ is dependent on the redshift ($z$) allowing the model to be conditioned to generate galaxies at a given redshift. Appendix \ref{appendix:model} shows the model architecture and details.

\begin{figure*}[tbh]
    \includegraphics[width=5.75in]{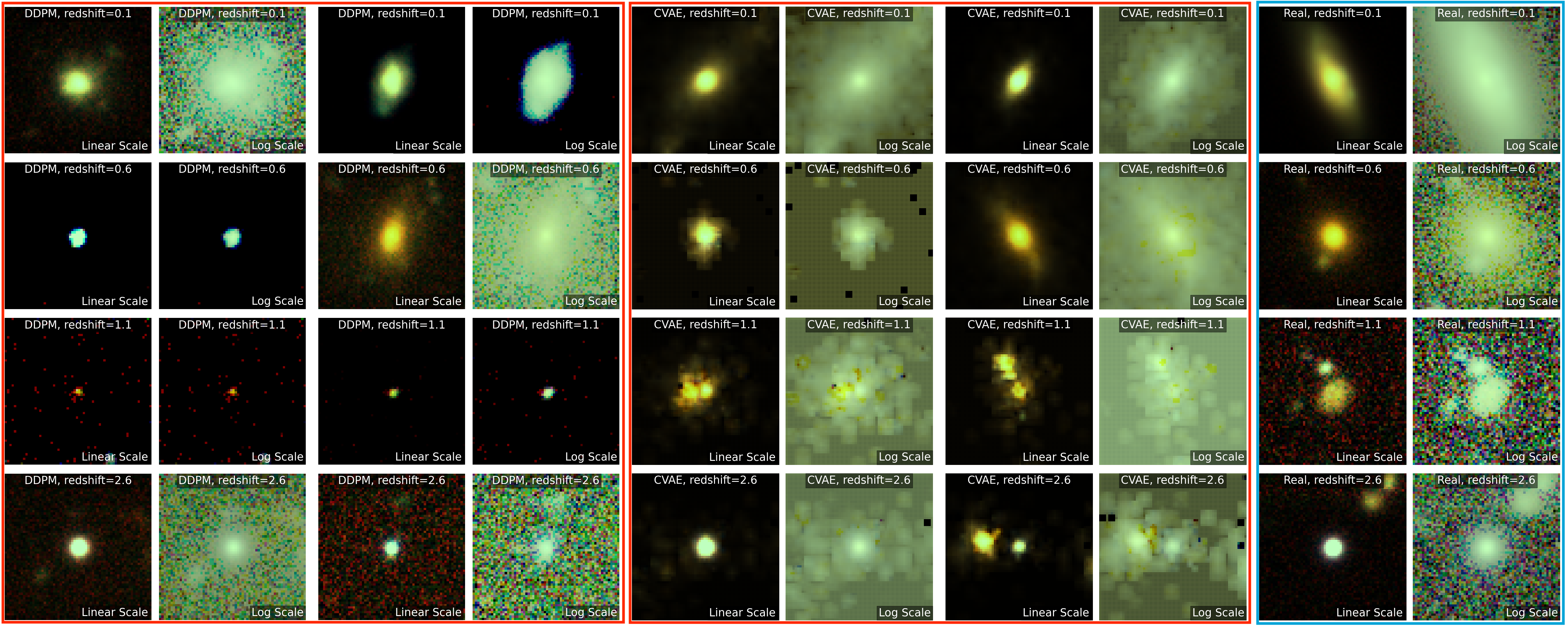}
    \caption{Examples of galaxy images that were generated by the DDPM (column 1 to 4) and the CVAE (columns 5 to 8) compared to real images (columns 9 to 10). Each row is at a different redshift. Each galaxy is shown with both linear and log image scaling. The galaxy images produced by the CVAE tend to have artifacts and correlated pixels in the background, which is clearly seen in log scale. The CVAE images of higher redshift ($>1$) galaxies also tend to show extended irregular bright artifacts around the central galaxy that is not in the real images.}
    \label{fig:generated_galaxies}
\end{figure*}

\subsubsection{Conditional Variational Autoencoder} \label{cvae}

Our CVAE uses a typical design with an encoder and a decoder, both of which are conditioned on the galaxy redshift. The encoder consist of a convolutional neural network (CNN) with the five mutli-wavelength images and the redshift of the galaxy as inputs to learn a latent probability distribution. This CNN uses techniques adopted from the HD-CNN \citep{yan2015}, where features extracted from different levels of convolution are concatenated. We condition the redshift along with these flattened feature maps, and then pass the combination through dense layers. The encoder produces a 64-variable normal distribution, which is the vector for the latent posterior distribution. The decoder takes the latent vector and the condition to output 5 images conditioned on the redshift. 

The decoder mirrors the encoder in terms of the dimensions at each convolutional step. The decoder inputs consists of the redshift and a latent vector sampled from the latent posterior distribution. The concatenated input is passed through dense layers, and then to an inverted HD-CNN architecture which produces the original image. Appendix X shows the model architecture and details.

We train the CVAE by of maximizing the variational lower bound as described in \citet{sohn2015}, expressed as: 
\begin{equation}
\log p_{\theta}(y|x) \geq -KL(q_{\phi}(z|x, y) || p_{\theta}(z|x)) + \mathbb{E}_{q_{\phi}(z|x, y)}[\log p_{\theta}(y|x, z)]
\end{equation}

In this expression, $x$ is the input condition, $z$ is the latent space representation, and $y$ is the output image [Sohn 2015]. The encoder output is $q_{\phi}(z|x, y)$, while the decoder output is $\log p_{\theta}(y|x, z)$. KL-Divergence regulates the latent distribution. For our model, we assume a Gaussian prior $p_{\theta}(z|x)$. The weighted average of the log likelihood in the second term is approximated in the random sampling process of decoder input. The loss function composes of the mean squared error between the input training image and the output plus the KL divergence regularization term. The mean squared error term and the regularization term are balanced using a KL weight hyperparameter, which we set to be $10^{-3}$ (The mean squared error term is $1000$ times stronger).

\section{Results}
The CVAE and the DDPM both generate images with visually realistic central galaxies, but the DDPM generates fewer artifacts and more realistic background properties. 
We generated $2,000$ images from the CVAE and $2,000$ images with the DDPM conditioned on redshift values drawn from a uniform distribution between 0.001 and 4. 
In Figure \ref{fig:generated_galaxies}, we show representative generated images at redshifts of 0.1, 0.6, 1.1, and 2.6, which correspond to lookback times of 1.32 Gyr, 5.85 Gyr, 8.29 Gyr, and 11.26 Gyr years ago (based on $\Lambda CDM$, the standard model of cosmology). 
Qualitatively, the generated galaxies are larger (larger isophotal area) at lower redshifts (more recent lookback time) and are more compact at higher redshifts when galaxies have had less time to evolve and interact. 
However, the CVAE tend to generate more extended structures around the central galaxy (see Fig. \ref{fig:generated_galaxies}) at high redshifts than the DDPM. These structures are not in real images of galaxies at high redshifts. In addition, the DDPM images have much more realistic background properties than the CVAE. 
The background in real astronomical images are the result of sky and detector noise, which generally have Gaussian or Poissonian distribution in pixel values. 
The background pixels in real images are also usually uncorrelated. 
The CVAE produces structure and correlated pixels in the background which are not present in the original images. This is more apparent when the image pixel values are displayed on a log scale (Fig. \ref{fig:generated_galaxies}). 
The DDPM is able to produce more realistic random background properties likely because the training process uses uncorrelated Gaussian noise.

\begin{figure}[htb]
\centering
    \includegraphics[width=5.50in]{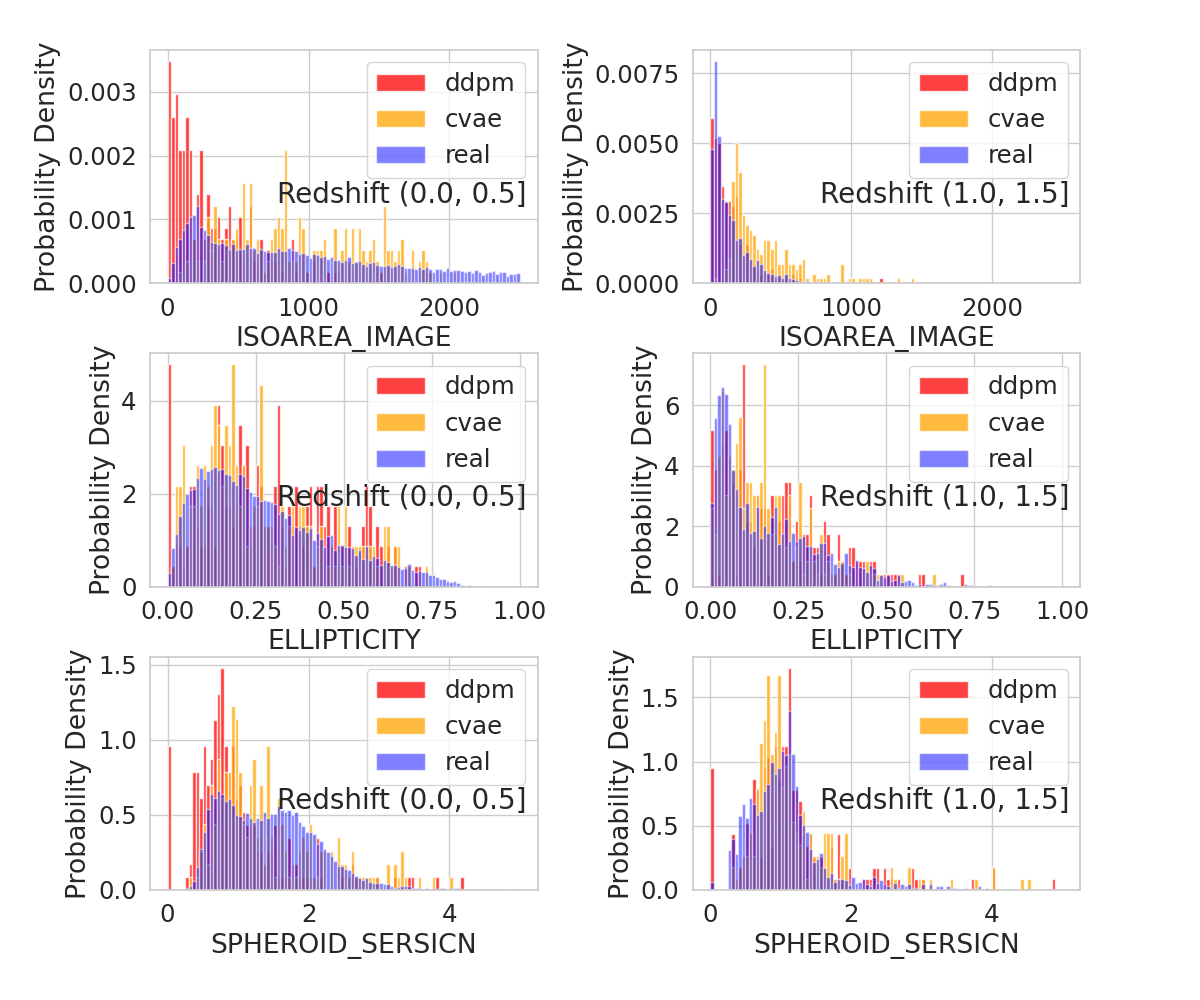}
    \caption{Distribution of the three physical parameters (top: isophotal area, middle: ellipticity, bottom: Sersic index at two different redshift bins (left: $<0.5$, right: $1.0 < z < 1.5$) for the real images (blue), DDPM images (DDPM), and CVAE images (orange). The CVAE distributions matches the real data at redshifts $<0.5$, but the DDPM matches better at higher redshifts.}
    \label{fig:parameter_distribution_detail}
\end{figure}

The quantitative galaxy metrics also generally agree with the qualitative visual assessment and contribute a more reliable and interpretable assessment of the models.
As a galaxy evolves over billions of years, it grows to become more diverse in shape and size as it interacts with its environment and other nearby galaxies. In addition, the light distribution and color of the galaxy also evolves with time as its stars form, grow old, and die. 
These characteristics of galaxy evolution result in a large dispersion in the distribution of isophotal area, ellipiticity, Sersic index for real galaxies at low redshifts (closer lookback time) and sharper distributions at high redshift (further lookback time) (Fig. \ref{fig:parameter_distribution_detail}). 
The DDPM is better than the CVAE at producing the distribution of physical properties of the galaxies for redshifts $>0.5$, but CVAE is better matched at redshift $<0.5$. 
At the lowest redshift bins (from 0 to 0.5), the DDPM images have smaller galaxies than those found in the real images. 
However, beyond redshift 0.5, compared to the CVAE, the DDPM produces galaxies with distributions of isophotal area, ellipiticity, Sersic index much closer to the real galaxies (Fig. \ref{fig:parameter_distribution_detail} \& \ref{fig:parameter_distribution}). Table \ref{tab:galaxy_metrics} summarizes the quantitative metrics averaged over the entire generated dataset. 

\begin{figure}[thb]
\centering
    \includegraphics[width=5.0in]{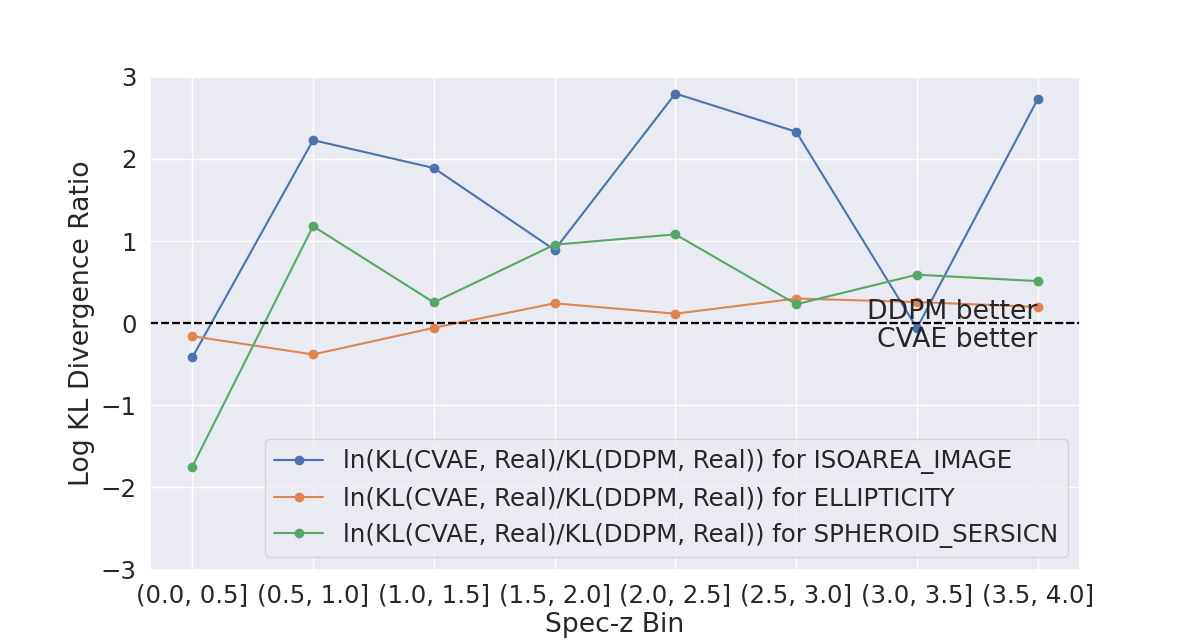}
    \caption{The log ratio of the KL Loss between the CVAE and the DDPM as a function of redshifts for isophoto area (blue), ellipticity (orange), and Sersic index (green). The CVAE performs better at the lowest redshift bin, but DDPM performs better over a larger range of redshifts.}
    \label{fig:galaxy_kl_loss}
\end{figure}

The KL divergence of the distribution of the real galaxy properties compared to the generated sample can be used to evaluate how well the models reproduce galaxy properties as a function of redshift (Fig. \ref{fig:galaxy_kl_loss}. 
The KL divergence for the distribution in isophotal area is a factor of 3 worse for the DDMP model compared to the CVAE for redshifts $< 0.5$. However, for redshifts $> 0.5$ the DDMP is a factor of 10 to 100 better than the CVAE at producing the distribution of isophotal area seen in real galaxies; the CVAE tends to produce larger galaxies than measured in real images. 
This is in line with our visual inspection of the images of high redshift galaxies generated by the CVAE, which show unusual extended structures around the central galaxy. The ellipticity and the Sersic index also show similar trends with redshift. We note however that while the CVAE is producing galaxies closer to the real data at low redshifts ($< 0.5$), both models cannot yet produce the true diversity seen in real galaxies for this range of redshift. At higher redshifts ($>0.5$), the DDPM produces galaxies with properties much closer to real galaxies (see Appendix \ref{appendix:comparison}). 

\begin{figure}[thb]
\centering
    \includegraphics[width=5.0in]{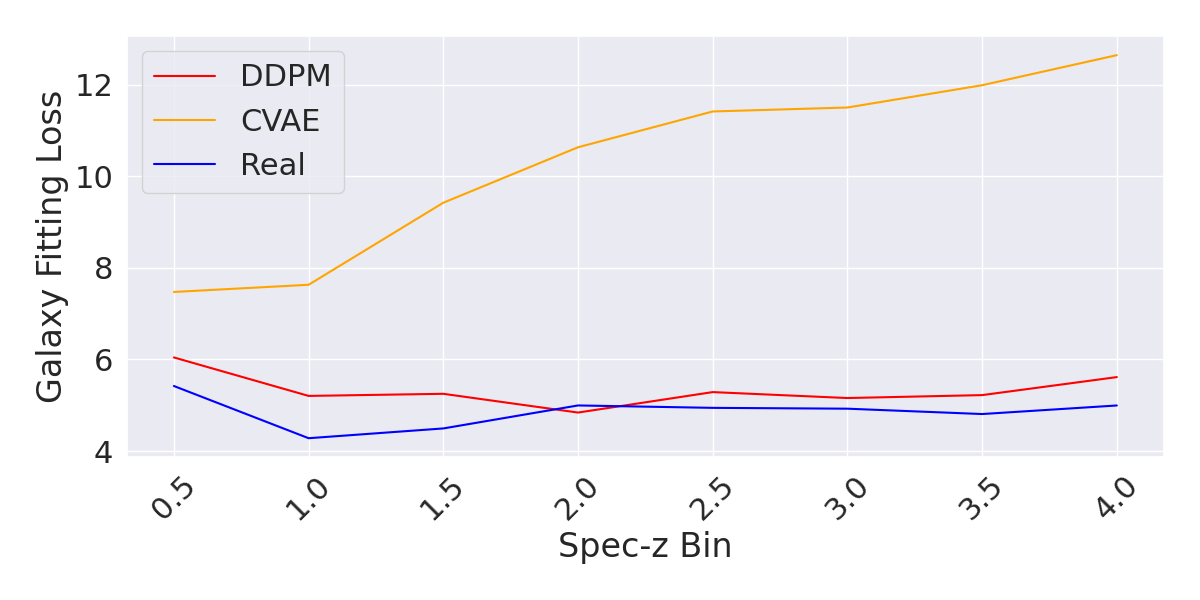}
    \caption{Galaxy fitting loss as a function of redshifts for the real data (blue), DDPM (red), and CVAE (orange). Note that the DDPM model can produce a lower fitting loss than real galaxies. This can be attributed to generative models' tendency to denoise. }
    \label{fig:galaxy_fitting_loss}
\end{figure}

The DDPM model outperforms the CVAE model in terms of the galaxy-fitting loss metric, as shown in Figure \ref{fig:galaxy_fitting_loss}. The DDPM model shows a galaxy-fitting loss close to the true dataset. 
This means that the DDPM generated galaxies share a closer level of irregularity and noise-to-signal ratio as real galaxy images than the CVAE images. 
The CVAE model has a high galaxy-fitting loss over all the ranges, although at a lower redshift the loss is smaller.

\begin{figure}[tbh]
\centering
    \includegraphics[width=5.0in]{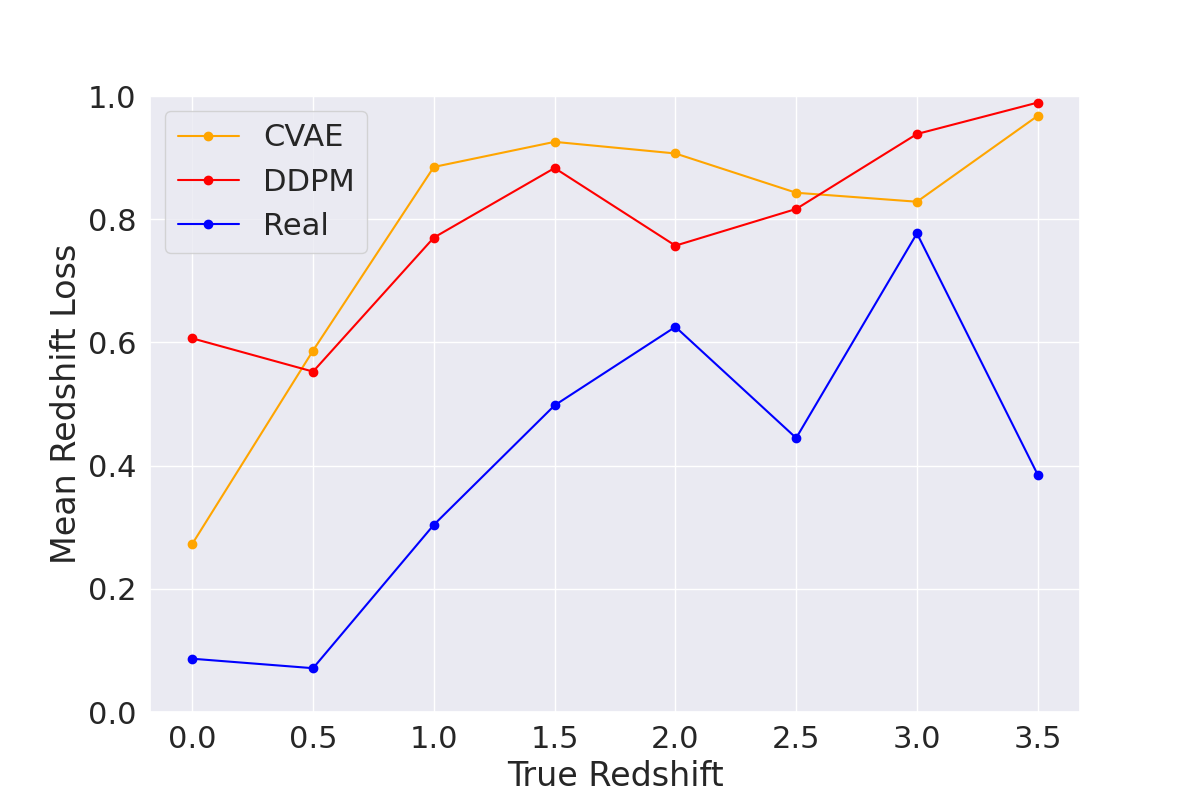}
    \caption{Redshift loss as a function of the true redshift for real images (blue), DDPM (red), and CVAE (orange). Neither models are able produce images that are consistent with the redshifts they are conditioned on based on the pretrained CNN model for predicting redshifts from real images.}
    \label{fig:redshift_loss}
\end{figure}

Both the DDPM and CVAE models perform poorly in the redshift loss metric. The galaxies generated by these models are not predictive of the redshift they were conditioned on. 
The difference between the conditioned redshift and the predicted redshift by the CNN model can be greater than 1 (which can represent billions of years of time), which is not useful for scientific applications. Fig. \ref{fig:redshift_loss} shows the redshift loss as a function of redshift while Appendix \ref{appendix:comparison} shows the predicted vs conditioned redshifts. 

We find that the IS and the FID metrics to be marginally helpful in quantifying whether the generated galaxies are realistic (Table \ref{tab:galaxy_metrics}). The absolute values for these metrics are not meaningful as the metrics were created for the ImageNet set of images, which are very disimilar to these galaxy images. The relative scores however, do indicate that the DDPM, on average, produces closer images to the real galaxies than the CVAE  (the DDPM IS score is higher than the CVAE and the FID score for the DDPM is lower than for the CVAE). This is the same conclusion that would be drawn from visual inspection. The quantitative metrics give a better characterization in the ways which the models differ in their abilities. 

\begin{table}[htb]
\centering
\begin{tabular}{lccc}
\hline
\textbf{Metric} & \textbf{Real} & \textbf{CVAE} & \textbf{DDPM} \\
\hline
Galaxy KL Loss & - & 0.23 & 0.22 \\
Galaxy Fitting Loss & 4.92 & 10.2 & 5.32 \\
Redshift Loss & 0.072 & 0.59 & 0.62 \\
IS & 1.72 & 1.55 & 1.87 \\
Fréchet Inception Distance & 1.70 & 123 & 98.8 \\
\hline
\end{tabular}
\caption{Comparison of metrics for real and generated galaxy images.}
\label{tab:galaxy_metrics}
\end{table}

\section{Discussion \& Conclusion}

As we have demonstrated using the CVAE and the DDPM models, galaxy images can be used as a physics-based ground truth for generative models, and galaxy metrics offer additional important ways to evaluate model performance. Using galaxy metrics allow us to quantitatively and objectively evaluate a generative model's capability. The structural properties of galaxies are simple enough that they can be modeled and quantified in both real and generated images. Importantly, the physics of galaxies are complex enough to produce enough diversity that it can challenge the ability of generative models to produce realistic galaxy images over time (redshift). 

While the CVAE and DDPM produce visually very similar galaxies based on human inspection, the physics based metrics show the limitations of these models. 
From galaxy metrics, the DDPM model performed better in reproducing physics based features at higher redshifts ($>0.5$), while the CVAE produced more realistic galaxies at lower redshifts ($<0.5$). 
However, both models were not able to produce images of galaxies that encode accurate redshift values. 
In addition, neither model can reproduce the diversity of galaxies seen at lower redshifts. 
This suggests that the models have a difficult time in encoding some of the physics of galaxy evolution that causes the appearance of galaxies to evolve over time. 

For this work, we chose metrics that easily evaluate measurements from galaxy images, and we aim to incorporate astronomers' discoveries of more complex relationships in future work. 
For example, most of the stars in the Universe are formed in the epoch of star formation which occurred between redshifts 1 to 3, peaking at redshift 2. 
Using additional imaging filters, it is possible to measure the star formation rate of galaxies. 
This type of relationship is more complex than reproducing structural features in the images of galaxies and can provide new challenges for generative models to learn physical relationships. 

\bibliography{AstroML}

\begin{thebibliography}{}
\expandafter\ifx\csname natexlab\endcsname\relax\def\natexlab#1{#1}\fi
\providecommand{\url}[1]{\href{#1}{#1}}
\providecommand{\dodoi}[1]{doi:~\href{http://doi.org/#1}{\nolinkurl{#1}}}
\providecommand{\doeprint}[1]{\href{http://ascl.net/#1}{\nolinkurl{http://ascl.net/#1}}}
\providecommand{\doarXiv}[1]{\href{https://arxiv.org/abs/#1}{\nolinkurl{https://arxiv.org/abs/#1}}}

\bibitem[{Aihara {et~al.}(2019)Aihara, AlSayyad, Ando, Armstrong, Bosch, Egami,
  Furusawa, Furusawa, Goulding, Harikane, Hikage, Ho, Hsieh, Huang, Ikeda,
  Imanishi, Ito, Iwata, Jaelani, Kakuma, Kawana, Kikuta, Kobayashi, Koike,
  Komiyama, Li, Liang, Lin, Luo, Lupton, Lust, MacArthur, Matsuoka, Mineo,
  Miyatake, Miyazaki, More, Murata, Namiki, Nishizawa, Oguri, Okabe, Okamoto,
  Okura, Ono, Onodera, Onoue, Osato, Ouchi, Shibuya, Strauss, Sugiyama, Suto,
  Takada, Takagi, Takata, Takita, Tanaka, Terai, Toba, Uchiyama, Utsumi, Wang,
  Wang, \& Yamada}]{aihara2019}
Aihara, H., AlSayyad, Y., Ando, M., {et~al.} 2019, Publications of the
  Astronomical Society of Japan, 71, 114, \dodoi{10.1093/pasj/psz103}

\bibitem[{Arjovsky {et~al.}(2017)Arjovsky, Chintala, \& Bottou}]{arjovsky2017}
Arjovsky, M., Chintala, S., \& Bottou, L. 2017, Wasserstein {{GAN}},
  https://arxiv.org/abs/1701.07875v3

\bibitem[{Barratt \& Sharma(2018)}]{barratt2018}
Barratt, S., \& Sharma, R. 2018, A {{Note}} on the {{Inception Score}},  arXiv.
\newblock \doarXiv{1801.01973}

\bibitem[{Bertin \& Arnouts(1996)}]{bertin1996}
Bertin, E., \& Arnouts, S. 1996, Astronomy and Astrophysics Supplement Series,
  117, 393, \dodoi{10.1051/aas:1996164}

\bibitem[{Chong \& Forsyth(2020)}]{chong2020}
Chong, M.~J., \& Forsyth, D. 2020, Effectively {{Unbiased FID}} and {{Inception
  Score}} and Where to Find Them,  arXiv, \dodoi{10.48550/arXiv.1911.07023}

\bibitem[{Conselice(2014)}]{conselice2014}
Conselice, C.~J. 2014, Annual Review of Astronomy and Astrophysics, 52, 291,
  \dodoi{10.1146/annurev-astro-081913-040037}

\bibitem[{Do {et~al.}(2024)Do, Jones, Boscoe, Li, \& Alfaro}]{do2024}
Do, T., Jones, E., Boscoe, B., Li, Y.~B., \& Alfaro, K. 2024, {{GalaxiesML}}:
  An Imaging and Photometric Dataset of Galaxies for Machine Learning,  Zenodo,
  \dodoi{10.5281/zenodo.11117528}

\bibitem[{Fussell \& Moews(2019)}]{fussell2019}
Fussell, L., \& Moews, B. 2019, Monthly Notices of the Royal Astronomical
  Society, 485, 3203, \dodoi{10.1093/mnras/stz602}

\bibitem[{Goodfellow {et~al.}(2014)Goodfellow, {Pouget-Abadie}, Mirza, Xu,
  {Warde-Farley}, Ozair, Courville, \& Bengio}]{goodfellow2014}
Goodfellow, I., {Pouget-Abadie}, J., Mirza, M., {et~al.} 2014, in Advances in
  {{Neural Information Processing Systems}} 27, ed. Z.~Ghahramani, M.~Welling,
  C.~Cortes, N.~D. Lawrence, \& K.~Q. Weinberger (Curran Associates, Inc.),
  2672--2680

\bibitem[{Heusel {et~al.}(2018)Heusel, Ramsauer, Unterthiner, Nessler, \&
  Hochreiter}]{heusel2018}
Heusel, M., Ramsauer, H., Unterthiner, T., Nessler, B., \& Hochreiter, S. 2018,
  {{GANs Trained}} by a {{Two Time-Scale Update Rule Converge}} to a {{Local
  Nash Equilibrium}},  arXiv, \dodoi{10.48550/arXiv.1706.08500}

\bibitem[{Ho {et~al.}(2020)Ho, Jain, \& Abbeel}]{ho2020}
Ho, J., Jain, A., \& Abbeel, P. 2020, Denoising {{Diffusion Probabilistic
  Models}},  arXiv, \dodoi{10.48550/arXiv.2006.11239}

\bibitem[{Jones {et~al.}(2024)Jones, Do, Boscoe, Singal, Wan, \&
  Nguyen}]{jones2024}
Jones, E., Do, T., Boscoe, B., {et~al.} 2024, The Astrophysical Journal, 964,
  130, \dodoi{10.3847/1538-4357/ad2070}

\bibitem[{Kingma \& Welling(2013)}]{kingma2013}
Kingma, D.~P., \& Welling, M. 2013, Auto-{{Encoding Variational Bayes}},
  https://arxiv.org/abs/1312.6114v11

\bibitem[{Krizhevsky {et~al.}(2012)Krizhevsky, Sutskever, \&
  Hinton}]{krizhevsky2012}
Krizhevsky, A., Sutskever, I., \& Hinton, G.~E. 2012, in Advances in {{Neural
  Information Processing Systems}}, Vol.~25 (Curran Associates, Inc.)

\bibitem[{Lanusse {et~al.}(2021)Lanusse, Mandelbaum, Ravanbakhsh, Li, Freeman,
  \& P{\'o}czos}]{lanusse2021}
Lanusse, F., Mandelbaum, R., Ravanbakhsh, S., {et~al.} 2021, Monthly Notices of
  the Royal Astronomical Society, 504, 5543, \dodoi{10.1093/mnras/stab1214}

\bibitem[{Lee {et~al.}(2022)Lee, Kim, Shin, Tan, Liu, Meng, Qin, Chen, Yoon, \&
  Liu}]{lee2022}
Lee, S.-g., Kim, H., Shin, C., {et~al.} 2022

\bibitem[{Mirza \& Osindero(2014)}]{mirza2014}
Mirza, M., \& Osindero, S. 2014, arXiv:1411.1784 [cs, stat].
\newblock \doarXiv{1411.1784}

\bibitem[{Nishizawa {et~al.}(2020)Nishizawa, Hsieh, Tanaka, \&
  Takata}]{nishizawa2020}
Nishizawa, A.~J., Hsieh, B.-C., Tanaka, M., \& Takata, T. 2020, Photometric
  {{Redshifts}} for the {{Hyper Suprime-Cam Subaru Strategic Program Data
  Release}} 2,  arXiv, \dodoi{10.48550/arXiv.2003.01511}

\bibitem[{Ravanbakhsh {et~al.}(2017)Ravanbakhsh, Schneider, \&
  Poczos}]{ravanbakhsh2017}
Ravanbakhsh, S., Schneider, J., \& Poczos, B. 2017, Equivariance {{Through
  Parameter-Sharing}},  arXiv.
\newblock \doarXiv{1702.08389}

\bibitem[{Robertson {et~al.}(2019)Robertson, Banerji, Brough, Davies, Ferguson,
  Hausen, Kaviraj, Newman, Schmidt, Tyson, \& Wechsler}]{robertson2019}
Robertson, B.~E., Banerji, M., Brough, S., {et~al.} 2019, Nature Reviews
  Physics, 1, 450, \dodoi{10.1038/s42254-019-0067-x}

\bibitem[{Ronneberger {et~al.}(2015)Ronneberger, Fischer, \&
  Brox}]{ronneberger2015}
Ronneberger, O., Fischer, P., \& Brox, T. 2015, U-{{Net}}: {{Convolutional
  Networks}} for {{Biomedical Image Segmentation}},  arXiv,
  \dodoi{10.48550/arXiv.1505.04597}

\bibitem[{Salimans {et~al.}(2016)Salimans, Goodfellow, Zaremba, Cheung,
  Radford, \& Chen}]{salimans2016}
Salimans, T., Goodfellow, I., Zaremba, W., {et~al.} 2016, arXiv:1606.03498
  [cs].
\newblock \doarXiv{1606.03498}

\bibitem[{Smith {et~al.}(2022)Smith, Geach, Jackson, Arora, Stone, \&
  Courteau}]{smith2022}
Smith, M.~J., Geach, J.~E., Jackson, R.~A., {et~al.} 2022, Monthly Notices of
  the Royal Astronomical Society, 511, 1808, \dodoi{10.1093/mnras/stac130}

\bibitem[{{Sohl-Dickstein} {et~al.}(2015){Sohl-Dickstein}, Weiss,
  Maheswaranathan, \& Ganguli}]{sohl-dickstein2015}
{Sohl-Dickstein}, J., Weiss, E.~A., Maheswaranathan, N., \& Ganguli, S. 2015,
  Deep {{Unsupervised Learning}} Using {{Nonequilibrium Thermodynamics}},
  arXiv.
\newblock \doarXiv{1503.03585}

\bibitem[{Sohn {et~al.}(2015)Sohn, Lee, \& Yan}]{sohn2015}
Sohn, K., Lee, H., \& Yan, X. 2015, in Advances in {{Neural Information
  Processing Systems}}, Vol.~28 (Curran Associates, Inc.)

\bibitem[{Song {et~al.}(2020)Song, Meng, \& Ermon}]{song2020}
Song, J., Meng, C., \& Ermon, S. 2020, Denoising {{Diffusion Implicit Models}},
  \dodoi{10.48550/arXiv.2010.02502}

\bibitem[{Vaswani {et~al.}(2017)Vaswani, Shazeer, Parmar, Uszkoreit, Jones,
  Gomez, Kaiser, \& Polosukhin}]{vaswani2017}
Vaswani, A., Shazeer, N., Parmar, N., {et~al.} 2017, in Advances in {{Neural
  Information Processing Systems}}, Vol.~30 (Curran Associates, Inc.)

\bibitem[{Wu \& Peek(2020)}]{wu2020}
Wu, J.~F., \& Peek, J. E.~G. 2020, arXiv:2009.12318 [astro-ph].
\newblock \doarXiv{2009.12318}

\bibitem[{Yan {et~al.}(2015)Yan, Zhang, Piramuthu, Jagadeesh, DeCoste, Di, \&
  Yu}]{yan2015}
Yan, Z., Zhang, H., Piramuthu, R., {et~al.} 2015, in 2015 {{IEEE International
  Conference}} on {{Computer Vision}} ({{ICCV}}) (Santiago, Chile: IEEE),
  2740--2748

\bibitem[{Yin {et~al.}(2022)Yin, Eisenstein, Finkbeiner, \&
  Protopapas}]{yin2022}
Yin, J.~E., Eisenstein, D.~J., Finkbeiner, D.~P., \& Protopapas, P. 2022,
  Publications of the Astronomical Society of the Pacific, 134, 044502,
  \dodoi{10.1088/1538-3873/ac5847}

\bibitem[{Zhang {et~al.}(2016)Zhang, Xu, Li, Zhang, Wang, Huang, \&
  Metaxas}]{zhang2016}
Zhang, H., Xu, T., Li, H., {et~al.} 2016, {{StackGAN}}: {{Text}} to
  {{Photo-realistic Image Synthesis}} with {{Stacked Generative Adversarial
  Networks}}, \dodoi{10.48550/arXiv.1612.03242}

\bibitem[{Zhou {et~al.}(2017)Zhou, Cai, Rong, Song, Ren, Zhang, Yu, \&
  Wang}]{zhou2017}
Zhou, Z., Cai, H., Rong, S., {et~al.} 2017, Activation {{Maximization
  Generative Adversarial Nets}}, https://arxiv.org/abs/1703.02000v9

\end{thebibliography}


\appendix

\section{Models}
\label{appendix:model}
\subsection{Conditional Denoising Diffusion Probabilistic Model}

As described in Section \ref{ddpm}, our conditional DDPM model is built by adding a condition mechanism to the DDPM framework developed by \citet{ho2020}. 

The forward process is non-conditional and identical to the mechanisms presented in \citet{ho2020}. We use a fixed linear noise schedule with $\beta_1=10^{-4}$ and $\beta_T=0.02$, for $T=1000$. For each timestep, we add noise to the image by $x_t = \sqrt{1 - \beta_t}x_{t-1} + \sqrt{\beta_t}\varepsilon_{t-1}; \text{ where } \varepsilon_{t-1} \sim \mathcal{N}(0, I)$. For un-normalized galaxy images, the large dynamic range is effectively reduced in the noising process, and at $T=1000$ we obtain a Gaussian random pixel distribution. 

As described by \citet{sohl-dickstein2015}, property of the Markov chain allows the direct computation of a timestep $x_t = \sqrt{\overline{\alpha_t}} x_0 + \sqrt{1 - \overline{\alpha_t}} \varepsilon$, with $\alpha_t = 1 - \beta_t$ and $\overline{\alpha_t} = \prod_{i=1}^T \alpha_i$. In the forward step, we do not recursive noise an image. 

The reverse process is done recursively, with each iteration we predict the probability distribution of $x_{t-1}$, given the current image $x_{t}$, timestep $t$ and redshift $z$. We assume this probability distribution is Gaussian, and has a standard deviation $\sigma_t = \sqrt{\beta_t}$ \citep{ho2020}. The probability distribution can therefore be expressed as: 

\begin{equation}
    p_\theta(x_{t-1} | x_t) = \mathcal{N}(x_{t-1}; \mu_\theta(x_t, t, z), \sigma_t)
\end{equation}

We are interested in knowing the mean value of the distribution $\mu_\theta(x_t, t)$. \citep{ho2020} derives from the forward process formula that 

\begin{equation}
    \mu_t = \frac{1}{\sqrt{\overline{\alpha_t}}} \left( x_t - \frac{1 - \alpha_t}{\sqrt{1 - \overline{\alpha_t}}} \varepsilon_{t} \right)
\end{equation}

Therefore, the problem becomes predicting the theoretical value of $\varepsilon_{t}$ as if the forward process is done recursively. Once we train a model to achieve this, we can carry out the conditional DDPM process and sample $x_{t-1}$ from the following distribution. This formula is also presented by \citet{lee2022} for a conditional DDPM model. 

\begin{equation}
    p(x_{t-1}) = \frac{1}{\sqrt{\alpha_t}} \left( x_t - \frac{1-\alpha_t}{\sqrt{1-\overline{\alpha_t}}} \varepsilon_\theta(x_t, t, z) \right) + N(0, \sigma_t),
\end{equation}

Our model predicts $\varepsilon_\theta(x_t, t, z)$, which should be as close as possible to $\varepsilon_{t}$. We employ a U-Net architecture for this task \citep{ronneberger2015}, and add conditioning mechanism to the U-Net using the approach of conditional GANs \citep{mirza2014}. Our U-Net has $5\times64\times64$ inputs with five feature map resolutions. We use standard practices such as employing transformer sinusoidal position embedding for the noise step (time) condition \citep{vaswani2017}, adding 4-head self-attention mechanism at each upsampling or downsampling block, or the ``midblock", and accompanying convolution layers throughout the network with group normalization layers \citep{wu2020}. 

\begin{figure}[h]
    \begin{center}
    \includegraphics[width=6.5in]{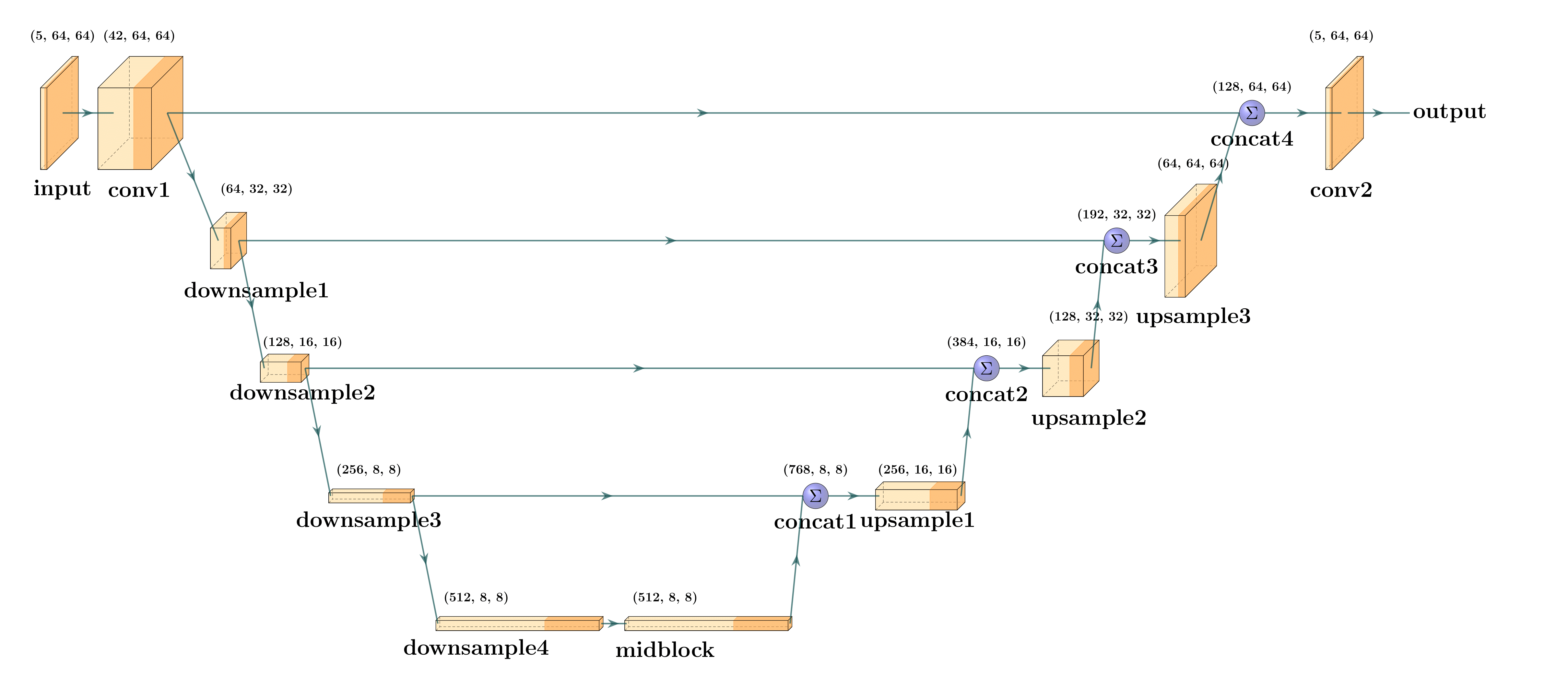}
    \end{center}
    \caption{Architecture of the U-Net we employed for the conditional DDPM model. }
    \label{fig:U-net}
\end{figure}

\begin{figure}[h]
    \begin{center}
    \includegraphics[width=6.5in]{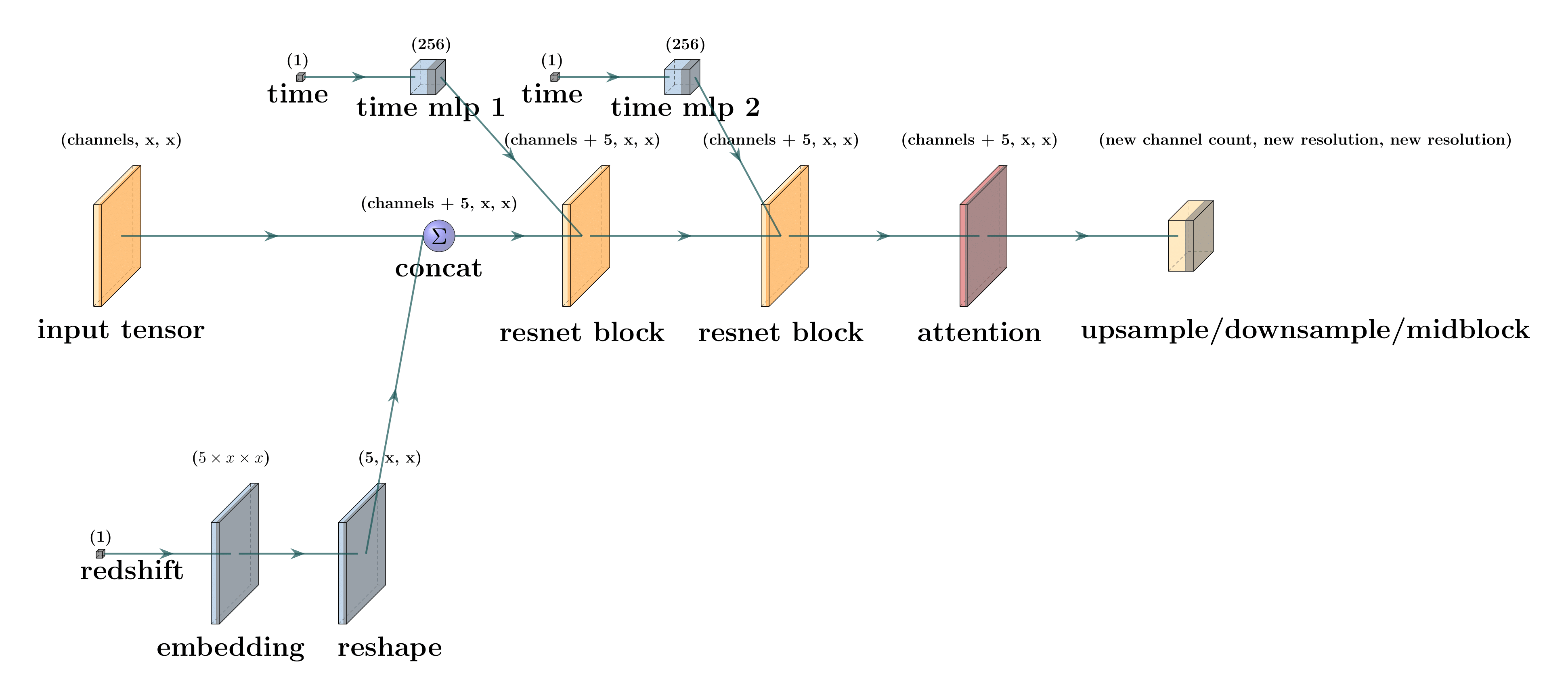}
    \end{center}
    \caption{Architecture of the upsampling/downsampling blocks. Within each ResNet block there are three convolutional layers, with Group Normalization and SiLu activation function. }
    \label{fig:block}
\end{figure}

Each upsampling or downsampling block, or the ``midblock" is added with the condition mechanism. Redshift label is converted to a hidden representation identical to the input of the blocks, then concatenated to these inputs. No embedding is done for the redshift condition value. A schematic of the model is shown in Figure \ref{fig:U-net} and \ref{fig:block}. 

Other hyperparameters include a batch size of $32$, a learning rate of $10^{-5}$. The resulting model has $1.24\times 10^8$ parameters, and is trained for $30$ epochs. 

\subsection{Conditional Variational Autoencoder}

\begin{figure}[h]
    \begin{center}
    \includegraphics[scale = 0.3]{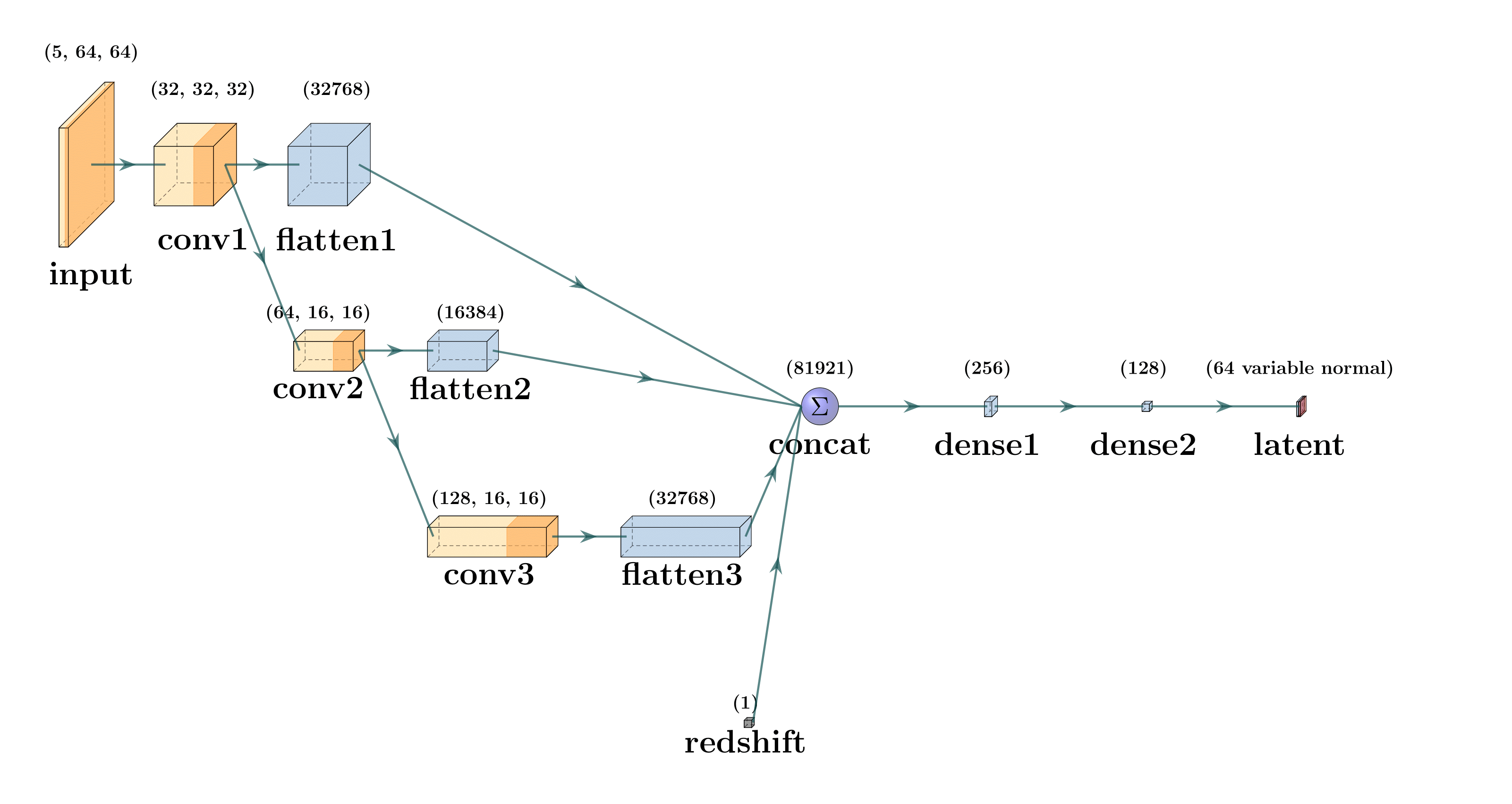}
    \end{center}
    \caption{Architecture of the encoder. }
    \label{fig:encoder}
\end{figure}

As described previously, our CVAE adopts a classic structure of an encoder and a decoder. We experimented with different variations and settled on a convolution layer arrangement similar to the one adopted in \citet{yan2015}, where each convolution layer in the sequence has its own skip layer to the final flatten layer. We found that this type of architecture give us the best balance between overall galaxy shape and fine details. Redshift condition is appended to the network via concatenation with the flattened feature maps. The feature map is connected to a series of dense layers. At the end, the encoder produces a 64-variable normal distribution, which is the vector for the latent posterior distribution. A schematic of the encoder is shown in Figure \ref{fig:encoder}. 

\begin{figure}[h]
    \begin{center}
    \includegraphics[scale = 0.3]{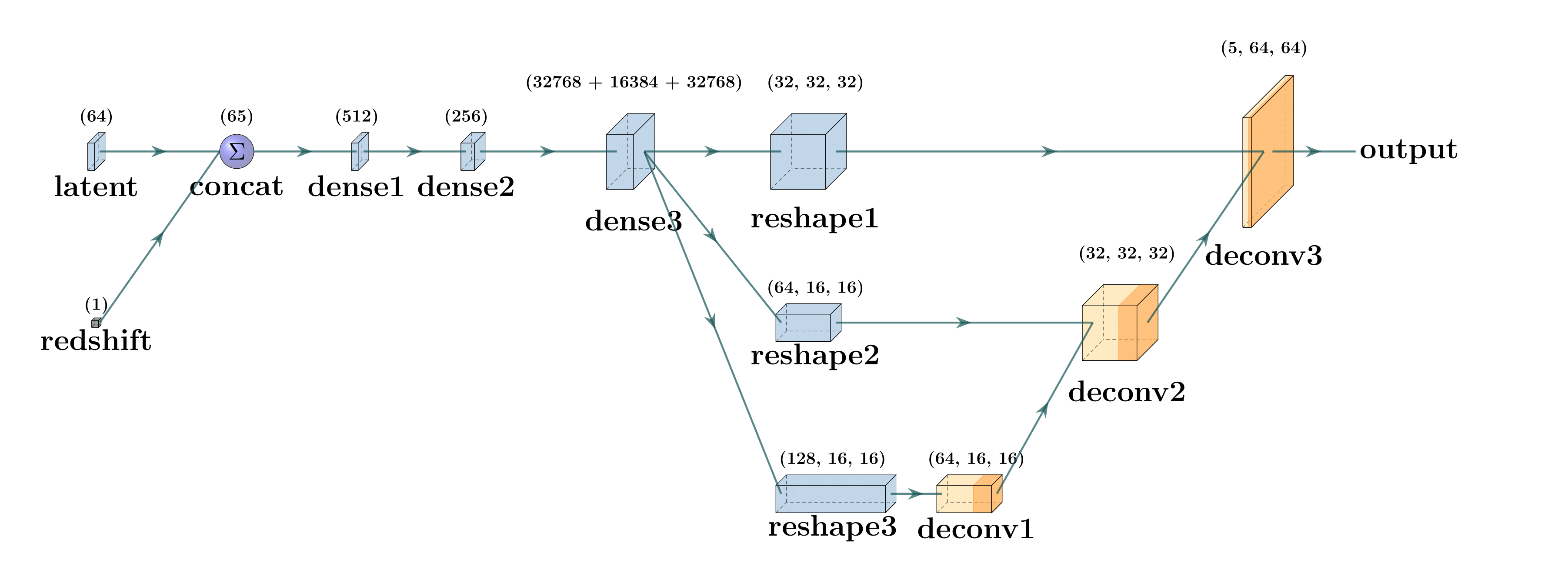}
    \end{center}
    \caption{Architecture of the decoder. }
    \label{fig:decoder}
\end{figure}

The decoder is symmetrical to the encoder with the convolution layers. However, because it does not need to account for the whole probability distribution of the latent space, and because redshift is processed differently, it has slightly less numbers of parameters as the encoder. A definite latent vector is sampled from the latent distribution as the decoder input. The decoder additionally takes redshift as its input. Both inputs are concatenated before passed through a series of dense and convolution layers with mirroring dimensions as the encoder. A schematic of the decoder is shown in Figure \ref{fig:decoder}. 

As mentioned in Section \ref{cvae}, the loss function is the sum of the mean squared error loss during regeneration and the regularization term. The regularization term measures the KL divergence between the latent space distribution and the prior. We assumed our prior to be a standard Gaussian, and use a weight of $10^{-3}$ on the regularization term. During training, we use a batch size of 512 to train for 250 epochs. 

\subsection{Redshift Discriminator - Convolutional Neural Network}

We use a modern convolutional neural network as our redshift discriminator \citep{krizhevsky2012}. The architecture is adopted from the redshift discriminator developed by Jones et al. (in prep.), which we modify the input dimension to $5\times64\times64$ to adapt to the shape of our input data. We found that not changing the other layers give the best prediction on real images for our data. This CNN has $4.85\times10^{6}$ parameters, and outputs a single point prediction of the redshift of the galaxy in the image.

\section{Evaluation Metrics for Generative Models based on Human Perception} \label{appendix:scores}

The IS (IS) and the Fréchet Inception Distance (FID) are efficient to calculate, and quantitatively are able to show the differences in model performances. In this work, we compare them with the metrics that we propose. 

In light of developments in generative models, especially GANs, \citet{salimans2016} argued that there is a lack of “objective analytical function” that evaluates generative models of images. They proposed the IS (IS) as an analytic metric that corresponds well to human perception. The IS is mathematically represented as: 
$$ IS(G) = \exp\left( \mathbb{E}_{\mathbf{x} \sim p_g} \left[ D_{KL}\left( p(y|\mathbf{x}) \, || \, p(y) \right) \right] \right) $$
The generated dataset is passed through the Inception model, which produces a probability distribution of the label for the input image. The KL divergence measures the difference between the probability distribution of labels over the whole dataset and the probability distribution of the labels evaluated on only the current input image $\mathbf{x}$. The average KL divergence across all generated data is calculated, and the IS is calculated as the exponent of this value. A high IS strongly indicates a generative model’s ability to produce images that are rated positively by human evaluators. 

The FID (Fréchet Inception Distance) is an improvement to the IS, which addresses the IS's limitation of not contrasting the statistics of real-world images with those of generated images. Introduced by \citet{heusel2018}, FID has been shown to be more sensitive to Gaussian disturbances compared to IS. The FID is calculated by first transforming both the original and generated datasets into feature vectors using the Inception model. The mean and covariance matrices of the real $(\mu_r, \Sigma_r)$ and generated $(\mu_g, \Sigma_g)$ feature vectors are then used to compute the FID: 

$$ FID = ||\mu_r - \mu_g||^2 + \mathrm{Tr}(\Sigma_r + \Sigma_g - 2(\Sigma_r \Sigma_g)^{1/2}) $$

Where $\mathrm{Tr}$ is the trace of the matrix. The formula assumes that the feature vectors follow multidimensional Gaussian distributions and calculates the Fréchet Distance between these distributions. 

Furthermore, several other methods have been suggested for assessing the quality of images, such as the AM score \citep{zhou2017} and the Wasserstein Critic \citep{arjovsky2017}. These metrics as well aim to either find an analytic value that correlates with human perception or to statistically compare the pixel values of generated and real images. However, it is questionable how objective a measurement that correlates well with human judgement means, especially for cases where humans cannot distinguish between real and generated images. To address this gap in evaluation methods, we propose our metrics based on galaxy image data. 

Despite their widespread successes, the difficulty to draw a direct line between human-evaluation based metrics and quantitative evaluations gives both metrics inherent limitations when used for scientific applications. \citet{chong2020} demonstrate that when the IS and FID are evaluated on finite datasets, the mean value, or expected value, do not align with their theoretical true values. Furthermore, the discrepancy between expected and true values is model-dependent, leading to potentially unreliable comparisons across different models. \citet{barratt2018} show that certain artificially constructed distributions, which differ strongly from real data, can yield higher IS than even the true distributions. They conclude that the IS and human perception are not always positively correlated, and also demonstrate that the IS can be misleading when applied to non-ImageNET based models. Such limitations prevent objective evaluation of generative models, when used scientifically, and require new metrics and evaluations to fill in the gap. 

\subsection{Additional Results}
\label{appendix:comparison}

We compute the performance the generated model images compared to real galaxy images as a function of redshift with respect to the isophotal area, ellipticity, and sersic index (Fig. 12).  Overall, the distribution of physical parameters produced by both generative models accross the entire redshift range successfully captures the distributions observed in the sample of real images.

We also calculate the KL divergence between the distribution of real and generated physical parameters for the CVAE and DDPM as a function of redshift (Fig. 13). The CVAE performs better with respect to the KL divergence across the three physical parameters at lower redshift bins compared to the DDPM ($z < 1.5)$, which performs better at higher redshifts ($z > 1.5)$. 

\begin{figure}[htb]
\centering
    \includegraphics[width=5.5in]{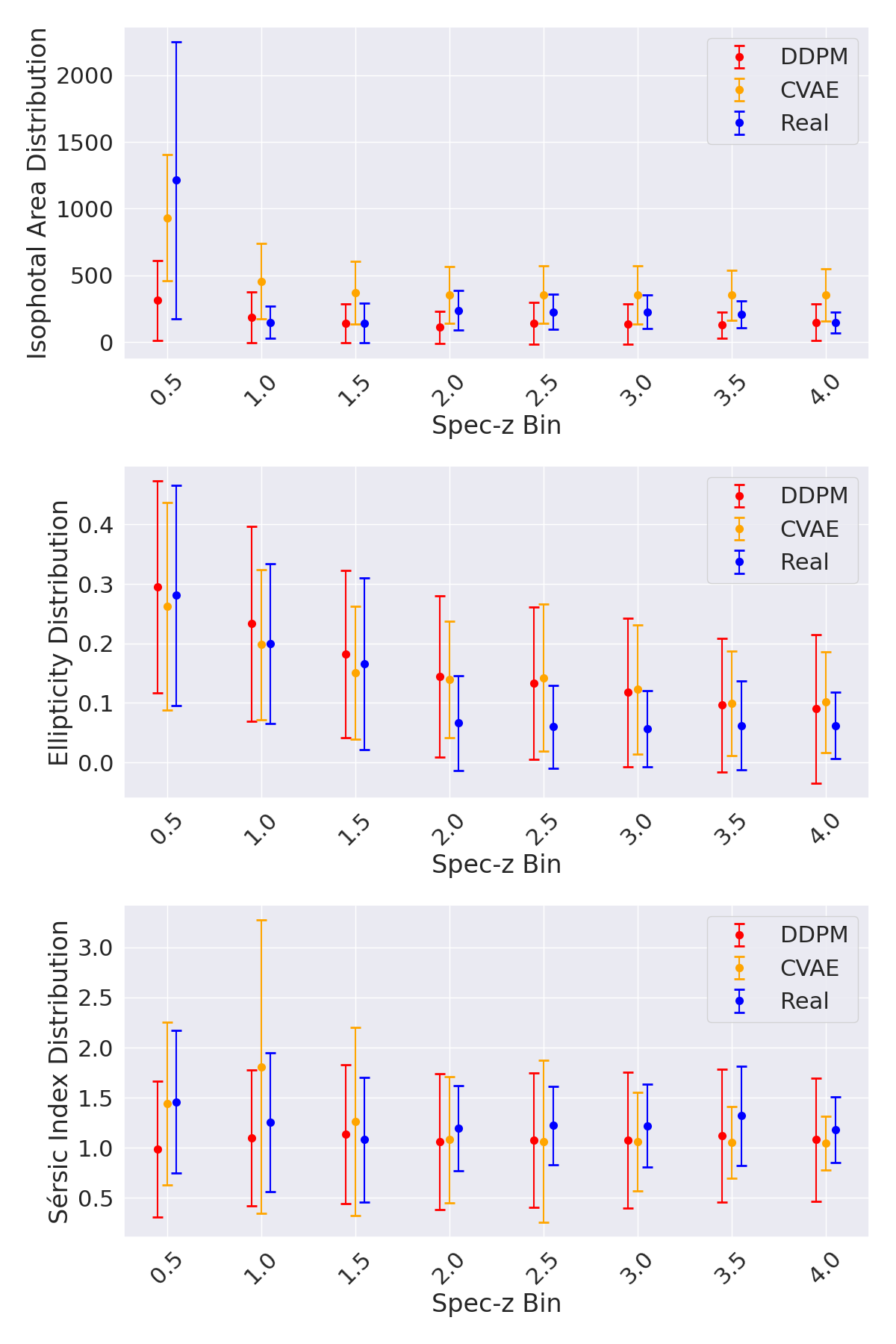}
    \caption{Mean and standard deviation of the isophotal area (top), ellipticity (middle), and Sersic index (bottom) for the real images (blue), DDPM images (red), and CVAE (blue) as a function of redshift. }
    \label{fig:parameter_distribution}
\end{figure}

\begin{figure}[thb]
\centering
    \includegraphics[width=5.0in]{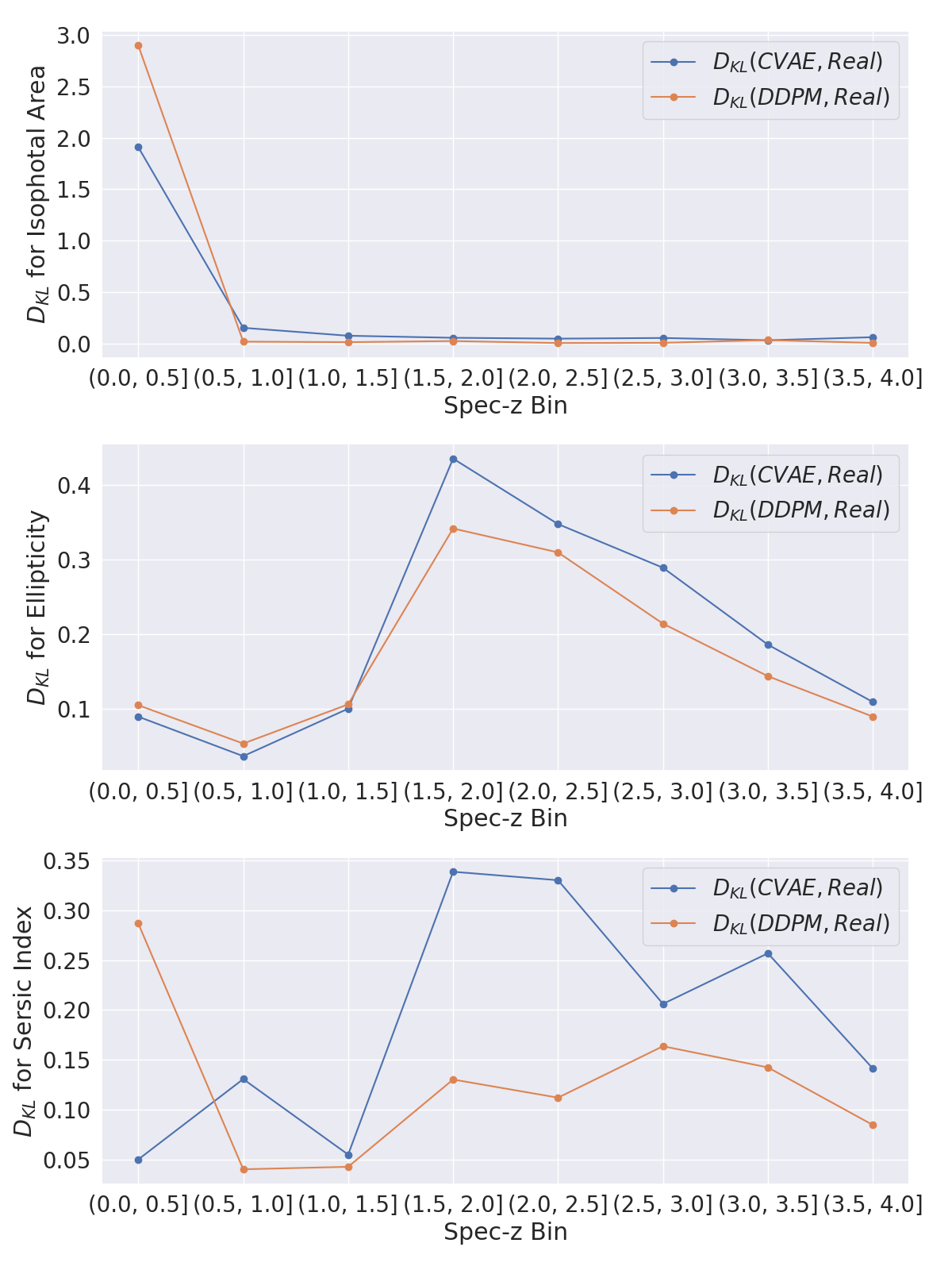}
    \caption{KL divergence between the real and generated (DDPM: red, CVAE: blue) for the isophotal area (top), ellipticity (middle), and Sersic index (bottom) as a function of redshift. }
    \label{fig:galaxy_kl_loss_detail}
\end{figure}

\begin{figure}[thb]
\centering
    \includegraphics[width=5.5in]{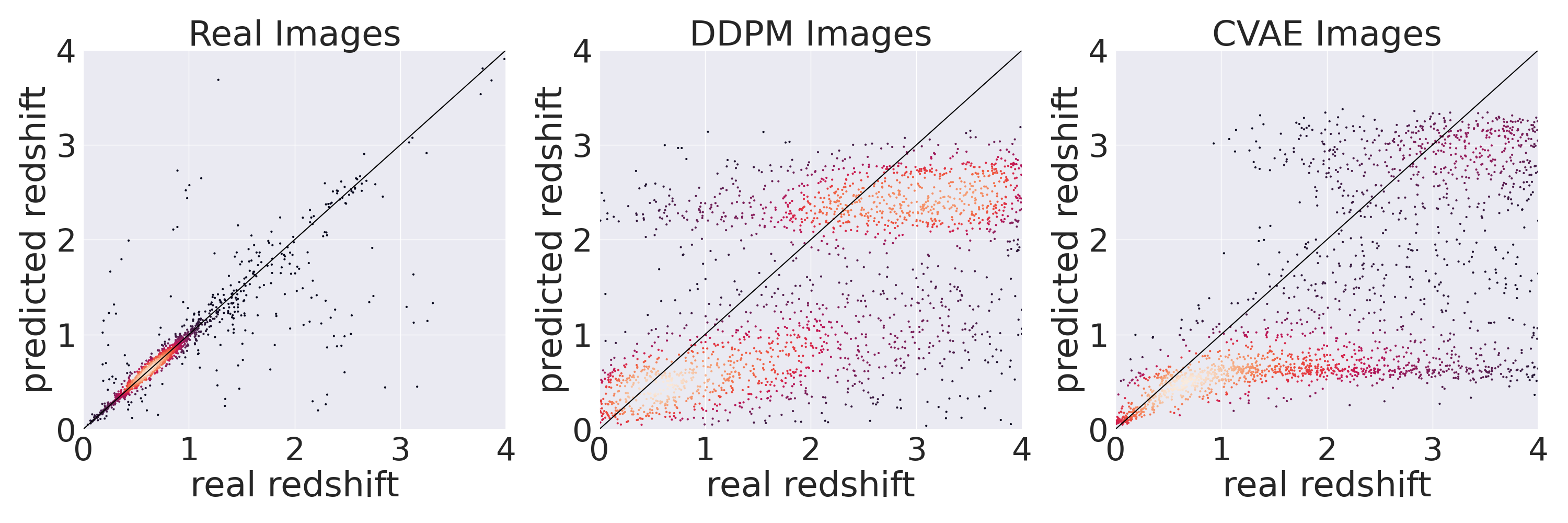}
    \caption{Comparison of the redshift predicted by the pretrained CNN to the true redshift for the real dataset (left), DDPM (center), and the CVAE (right). The line indicates where accurate redshift predictions should be. The color of the points indicate the density. }
    \label{fig:redshift_prediction_plots}
\end{figure}

\end{document}